\documentclass[12pt]{article}

\usepackage{graphics} 
\usepackage{blindtext}
\usepackage{epsfig} 
\usepackage{subfigure}
\usepackage{amsmath} 
\usepackage{amssymb}  
\usepackage{color}
\usepackage{lipsum}
\usepackage{cuted}
\usepackage{enumerate}
\newcommand{\norm}[1]{\left\lVert#1\right\rVert}

\title{\LARGE \bf
Identification of Markov Jump Autoregressive Processes \\ from Large Noisy Data 
Sets
}

\author{Sarah Hojjatinia$^{1}$,  Constantino M. Lagoa$^{2}$ 
\thanks{$^{1}$Sarah Hojjatinia is with the School of Electrical Engineering and Computer Science,
        The Pennsylvania State University, University Park, PA, USA,
        {\tt\small szh199@psu.edu}}%
    \thanks{$^{2}$Constantino M. Lagoa is with  the School of Electrical Engineering and Computer Science,
    	The Pennsylvania State University, University Park, PA, USA,
    	{\tt\small lagoa@psu.edu}
    	\newline
\indent
This work was partially supported by National Institutes of Health (NIH) Grant R01 HL142732, and National Science Foundation (NSF) Grant \#1808266.}
}

\date{}

\newtheorem{assumption}{Assumption} 
\newtheorem{problem}{Problem} 
\newtheorem{thm}{Theorem}

\newtheorem{Remark}{Remark}
\begin{document}

\maketitle
\thispagestyle{empty}
\pagestyle{empty}

\begin{abstract}
This paper introduces a novel methodology for the identification of switching 
dynamics for switched autoregressive linear models. Switching 
behavior is assumed to follow a Markov model. The system's outputs are  
contaminated by possibly large values of measurement noise. Although the 
procedure provided can handle other noise distributions, for simplicity, it is 
assumed that the distribution is Normal with unknown variance. 
Given noisy input-output data, 
we aim at identifying switched system coefficients, parameters of the noise distribution, dynamics of switching and probability transition matrix of Markovian model.	
System dynamics are estimated  using previous results which exploit algebraic 
constraints that system trajectories have to satisfy. Switching dynamics are 
computed with solving a maximum likelihood estimation problem.
The efficiency of proposed approach is shown with several academic examples. 
Although the noise to output ratio can be high, the method is shown to be 
extremely effective in the situations where a large number of measurements is 
available.
\end{abstract}

\section{Introduction}

While identification of linear time invariant systems is by now a well understood problem, identification of switched and hybrid systems is considerably less developed, even in the piecewise affine case. Existing methods exploit a number of algebraic, optimization-based techniques  to find subsystem dynamics and switching surfaces \cite{paoletti2007identification}. A common feature is the computational complexity entailed
in dealing with noisy measurements: in this case algebraic procedures lead to nonconvex optimization problems, while optimization methods lead to mixed integer/linear programming \cite{roll2004identification}.

Similarly, methods relying on probabilistic priors \cite{juloski2005bayesian}  lead to combinatorial problems. This can be avoided by using clustering-based methods \cite{nakada2005identification}. However, these require “fair sampling” of each cluster, which constrains the data that can be used. In \cite{ozay2012sparsification, Ozay2015180}, some sparsification based-techniques for identification of affine switched models have been developed that allow for several types of noise.


This paper develops effective methods for identifying switching dynamics  from 
large noisy
data sets, for a broad class of systems described by switching autoregressive 
models. These systems can be considered a generalization of piecewise affine 
models, and breach the gap between linear and nonlinear models, retaining many 
of the tractability properties of the former, while providing descriptions that 
more accurately capture the features of practical problems over broader 
scenarios.

Development of the proposed framework is motivated by health care applications: Specifically, smartphone-based  interventions for increasing light physical activity \cite{Lagoa2017}. 
More precisely, the availability of  activity tracking devices allows gathering of large
amount of data such as  physical activity of an individual. Physical activity is a dynamic behavior, so it can be  modeled as a dynamical system. Furthermore, its characteristics may remarkably change based on the time in a day, weekdays or weekends,  location, etc; which,  motivated the approach of modeling it as a switching system  \cite{conroy2019}.

In identifying the parameters of switched models, the dynamics of switching 
play an important role. The interest in Markovian jump systems, switched 
system with switching dynamics based on a Markov chain, has been growing since 
they have a broad range of application in different areas and real world 
problems such as economic systems, power systems, and networked control 
systems.  In comparison to the large amount of literature on analysis and 
control of Markovian jump systems, the identification problem seems to have 
received very little attention.

In \cite{HOJJATINIA201714088} a new method for the identification of parameters 
of Markovian jump system is provided. The probability transition matrix is 
estimated using a suitable convex optimization problem. However, due to 
computational complexity, the number of measurement that the proposed approach 
is able to handle is limited and only process noise was considered. 
In this paper, we focus on cases involving a very large number of measurements, 
possibly affected by large values of noise. In this case, polynomial/moments 
based approaches become
ineffective, and different methodologies need to be devised. The approach we 
propose builds upon the same premises as~\cite{hojjatinia2018identification}.

More precisely we start by assuming that 
the output measurements are corrupted  by random Normal measurement noise with 
unknown variance. Then, we exploit the availability of a large number of 
measurements and the results in~\cite{hojjatinia2018identification} to 
determine high confidence estimates of the systems parameters and the variance 
of the measurement noise.
Finally, by using a maximum likelihood approach, we estimate the probability 
transition matrix and dynamics of switching.  The approach can be easily 
extended to other noise distributions, as long as the number of unknown 
parameters of the distribution is "low."

 \subsection{Paper Organization}
The paper is structured as follows:   
after this introduction, problem statement is defined in Section \ref{stat}. Identification of system coefficients and noise parameters are reviewed in Section \ref{review}.
In Section \ref{ptm}, the method for identification of probability transition matrix is described. Numerical results are shown in Section \ref{result}.
Finally, Section \ref{conclusion} concludes the paper highlighting some possible future research directions.

 \section{Problem Statement}\label{stat}
A precise description of the problem addressed is provided in this section. 
Assumptions needed to solve the problem are also introduced.
 \subsection{System Model}
We consider switched autoregressive   (SAR) linear models  of the form
  \begin{align}
  x_{k}=\sum_{j=1}^{n_{a}}{a_{j  \delta_k}} \; x_{k-j}+\sum_{j=1}^{n_{c}}{c_{j \delta_k}} \;u_{k-j}    \label{eq:C}
  \end{align}
where $x_{k} \in \mathbb{R}$ is the output at time $k$ and $u_{k} \in \mathbb{R}$ is input at time $k$. The variable $\delta_k \in \{1, . . . , n\}$ denotes the sub-system active at time $k$, where  $n$ is the total number of sub-systems. Furthermore,  $a_{j  \delta_k}$ and $c_{j  \delta_k}$ denote unknown coefficients corresponding to mode $\delta_k$. Time $k$ takes values over the non-negative integers.
The latent discrete state $\delta_k$ evolves according to a Markov chain with 
transition probability matrix $P$, whose $ij$ entry is 
  \begin{align}
  P_{ij} = P \{ \delta_{k+1} = j ~|~  \delta_k  = i \} \label{eq:P}
  \end{align}
Output is assumed to be contaminated by (possibly large) noise; i.e.  
observations are of the form:
  \begin{align}
  y_{k}= x_{k}+\eta_{k}     \label{eq:C1}
  \end{align}
where $\eta_{k}$, denotes measurement noise. 

The following assumptions are made on the system model and noise.

\begin{assumption} \label{ass} 
Throughout this paper it is assumed that:
\begin{itemize} 
	\item  Upper bounds on $n_a$ and $n_c$ are available.
	\item  Upper bound on the number of sub-systems $n$ is available.
	\item  Measurement noise $\eta_{k}$ has zero mean Normal distribution with 
	unknown variance.
	\item  Noise $\eta_k$ is independent from $\eta_{l}$ for $k\neq l$, and identically distributed.   
	\item  Input sequence $u_{k}$  applied to the system is known and bounded. 
	\item  There exists a finite constant $L$ so that  $|x_{k}|\leq L$ for all positive integers $k$. 
	\item  Switching sequence is based on a Markov process.
\end{itemize}  
\end{assumption}  
\vskip 0.1in
Note that, the approach  in this paper can be extended for any noise 
distribution as long as the number of unknown distribution parameters is 
``small.''
\subsection{Problem Definition}
The main objective of this paper is to develop algorithms to  identify the 
parameters of SAR systems, noise parameters,   and  dynamics of switching from 
noisy observations. More precisely, we aim at solving the following problem:
\vskip 0.1in
\begin{problem}
Given Assumption~\ref{ass}, an input sequence $u_k$, $k=-n_c+1, \dots, N-1$ and noisy output measurements $y_k$, $k=-n_a+1, \dots, N$,
determine 
\begin{enumerate} 
	\item  Coefficients of the SAR model $a_{i,j}$, $i=1,2,\ldots, n_a$, $j=1,2,\ldots,n$, $c_{i,j}$, $i=1,2,\ldots, n_c$, $j=1,2,\ldots,n$,
	\item  Noise distribution parameters,
	\item  Switching sequence $\delta_k$ , $k=1,2,\ldots, n$ which is based on a Markov process.
\end{enumerate}  
\end{problem}
\vskip 0.1in
\section{Review: Identification of System Coefficients and Noise Parameters}  \label{review}
To identify the coefficients of SAR system with measurement noise from large amount of data, we adopt the approach developed in~\cite{hojjatinia2018identification}. For the sake of completeness, we briefly summarize the approach in this section. We refer the reader
to~\cite{hojjatinia2018identification} for more details.

First, we review earlier results on an algebraic reformulation of  the SAR identification problem for the case where no noise is present. Details on the algebraic approach to switched system identification can be found in~\cite{VidalSoattoMaEtAl2003}.

The equation~(\ref{eq:C}) is equivalent to
\begin{align}
b_{\delta_k}^{T} \; r_{k}=0  \label{eq:D}
\end{align}
where
\[
r_{k}=[x_{k},~ x_{k-1}, ~\cdots, ~x_{k-n_{a}}, ~u_{k-1}, ~\cdots, ~u_{k-{n_{c}}}]^{T}
\]
is the known regressor at time $k$,
and 
\begin{multline*}
{b_{\delta_k}=} ~ [-1,~a_{1 \delta_k}, ~\cdots, ~a_{n_{a} \delta_k},~c_{1 \delta_k}, ~\cdots, ~c_{n_{c} \delta_k}]^{T}.
\end{multline*}
is the vector of unknown coefficients at time $k$.
Hence, independently of which of the~$n$ submodels is active at time $k$, we have
\begin{align}
\Upsilon_{n}(r_{k})= \prod_{i=1}^{n} {b_{i}^{T} r_{k}}=c_{n}^{T} \nu_{n}(r_{k})=0,  \label{eq:E}
\end{align}
where the
vector of parameters corresponding to the $i$-th submodel is denoted by $b_{i} 
\in \mathbb{R}^{n_{a}+n_{c}+1}$,  and $\nu_{n}(\cdot)$ is  Veronese map of degree 
$n$ \cite{harris2013algebraic} 
\[
\nu_{n}([x_{1},~ \cdots,~  x_{s}]^{T} ) = [\cdots,~ x_{1}^{n_{1}}x_{2}^{n_{2}} \cdots x_{s}^{n_{s}} ,~\cdots]^{T}
\]
which contains all monomials of order $n$ in lexicographical order,
and $c_{n}$ is a vector whose entries are polynomial functions of unknown parameters $b_{i}$  (see \cite{vidal2005generalized}
for explicit definition). 
The Veronese map above is also known as  polynomial embedding in machine learning \cite{vidal2005generalized}. 
%

Note that the number of rows of the Veronese matrix~$V_n$ is equal to the 
number of measurements available for the regressor $N$. Therefore, a 
reformulation of the previous results  to address the problem of identification 
from very large data sets is as follows~\cite{hojjatinia2018identification}.
%

For the noiseless case, a reformulation of the hybrid decoupling constraint shows identifying the coefficients of the sub-models is equivalent to finding the singular vector $c_n$ associated with the minimum singular value of the matrix
\begin{equation} \label{eq:Mk}
\overline{\mathcal{M}}_N=\frac{1}{N} \sum_{k=1}^N \nu_{n}(r_{k}) \nu_{n}^T(r_{k}) \doteq \frac{1}{N} \sum_{k=1}^N M_k
\end{equation}
where, matrices are of size  $\binom{n+n_{a}+n_{c}}{n}$, and size does not depend on the number of measurements, which is  especially important in the case of very large data sets.


Identifying the parameters of the SAR model is equivalent to finding a vector 
in the null space of the matrix
$\overline{\mathcal{M}}_N$.
Under mild conditions, the null space of the matrix above has dimension one if 
and only if the data is compatible with the assumed model. However, when output 
is corrupted by noise, $x_k$ is not known and, therefore, this matrix cannot 
be  computed. 
However, we can use  available information on the statistics of the noise and 
the measurements collected to compute approximations of the matrix 
$\overline{\mathcal{M}}_N$ and,  consequently, approximations of vectors in its 
null space.

We start by noting that although $x_k$  are unknown, the following holds
	\begin{align} 
	E[x_{k}^{h}] &=E[(y_{k}-\eta_{k})^{h}] =E[y_{k}^{h}] - \sum_{d=1}^{h} \binom{h}{d}\, E[x_{k}^{h-d}]\, E[\eta_{k}^{d}] \nonumber \\
	&=E[y_{k}^{h}] - \sum_{d=1}^{h} \binom{h}{d}\, E[x_{k}^{h-d}]\, m_{d} \nonumber \\
	&\forall k=1, \, 2,\, \cdots, \, N. \label{eq:rec}
	\end{align} \vskip .2in
\noindent
where $E(\cdot)$ denotes expectation and $m_d$ is the $d^{th}$  moment of noise.

Hence, assuming that  distribution  of the noise, and the input signal are 
given and fixed, there exists an affine function $M(\cdot)$ so that
	\begin{align*}
	M_k &= E\{M[mon_n(y_k,\ldots,y_{k-n_a})]\}\\& = M\{E[mon_n(y_k,\ldots,y_{k-n_a})]\}.
	\end{align*}
where $mon_n(\cdot)$ denote a function that returns a vector with all monomials up to order $n$ of its argument. 
\vskip .2in

This can be exploited to identify the parameters of the SAR system. The only 
thing needed is an estimation of the  matrix~$\overline{\mathcal{M}}_N$ 
in~\eqref{eq:Mk}. It turns out that this can be done using the available noisy 
measurements. More precisely, we can construct the matrix
\[
\widehat{\overline{\mathcal{M}}}_N \doteq \frac{1}{N} \sum_{k=1}^N M[mon_n(y_k, 
\ldots, y_{k-n_a})]
\]
and it is shown in~\cite{hojjatinia2018identification} that this matrix 
converges to $\overline{\mathcal{M}}_N$ in~\eqref{eq:Mk} as $N\rightarrow 
\infty$ almost surely. Hence,  for large number of measurements $N$, the null 
space of the matrix $\widehat{\overline{\mathcal{M}}}_N$ can be used to 
determine the coefficients of the subsystems.
%

The above assumes knowledge of the moments of the noise. However, this does not 
need to be the case. For simplicity of exposition, assume that measurement 
noise has a Normal distribution with zero mean and unknown variance~$\sigma^2$. 
Then, since the moments are known functions of the variance,  
$\overline{\mathcal{M}}_N$ is a known function of $\sigma$ and  estimation of 
variance can be performed by minimizing the minimum singular value of matrix 
above over the allowable values of~$\sigma$.  More precisely, the parameters of 
the submodels and the variance of the noise can be identified using the following 
algorithm:
Let $n_a$, $n_c$, $n$, some parameters of the noise  and $\sigma_{\max}$  be 
given. 
\vskip .1in 
\begin{enumerate}[Step 1.] 
	\item Compute matrix
	$
	\widehat{\overline{\mathcal{M}}}_N
	$
	as a function of the unknown noise parameter $\sigma$.
	\item  Find the value  $\sigma^* \in [0,~\sigma_{\max}]$ that minimizes the minimum singular value of $\widehat{\overline{\mathcal{M}}}_N$.
	\item Let $c_n$ be associated singular vector.
	\item Determine the coefficients of the subsystems from the vector~$c_n$.
\end{enumerate}
\vskip 0.05in 
In order to perform Step 3 in Algorithm, we adopt 
polynomial differention algorithm for mixtures of hyperplanes, 
introduced by Vidal \cite[pp.~69--70]{vidal2003generalized}. 
%
In practice for sufficiently large $N$, the above algorithm provides both a good estimate of the systems coefficients and noise parameters, especially if we take $\sigma^*$ to be the smallest value of $\sigma$ for which the minimum singular value of $\widehat{\overline{\mathcal{M}}}_N$ is below a given threshold $\epsilon.$  
Previous work  cannot address the identification of switching dynamics and estimating the probability transition matrix of Markov jump models, this problem is explicitly  addressed in the following section.

\section{Identification of Probability Transition Matrix} \label{ptm}
In Section \ref{review}, the algorithms and procedure of identifying noise 
parameters  and system coefficients have been presented. 
In this section the switching behavior and dynamics of switching are 
considered. This is done in two steps: The first step is to identify switches 
that have the 
highest probability of occurrence. Then, in the second step, by considering 
these switches as a 
good estimate of switching sequence,  we estimate the  
transition probabilities.

\vskip .1in 
\subsection{ Maximum likelihood switch sequence}
Assume that the  noise variance and system coefficients have been identified. 
To do the first step in identification of switching dynamics, i.e., determine the 
switches with highest probability, we start by building the following sequence 
based on available data and identified coefficients and parameters. 
Considering  equations \eqref{eq:C} and \eqref{eq:C1}:
  \begin{align*}
  &x_{k}=\sum_{j=1}^{n_{a}}{a_{j  \delta_k}} \; x_{k-j}+\sum_{j=1}^{n_{c}}{c_{j \delta_k}} \;u_{k-j}  
\\
&  y_{k}= x_{k}+\eta_{k}     
  \end{align*}
Since $x_{k}= y_{k}-\eta_{k}  $.
%
we have  
\begin{equation} \label{eq:pn1}
\begin{aligned}
 \eta_{k} - &\sum_{j=1}^{n_{a}}{a_{j  \delta_k}} \;  \eta_{k-j} = \\
&y_{k} - \sum_{j=1}^{n_{a}}{a_{j  \delta_k}} \; y_{k-j} - \sum_{j=1}^{n_{c}}{c_{j\delta_k}} \;u_{k-j} 
\end{aligned}
\end{equation}
\noindent
Since we have identified the coefficients ${a_{j  \delta_k}}$ and 
${c_{j\delta_k}}$, and input output ($u$, $y$) are available, we are able to 
determine the realization of the random variable in the right hand side of 
equation (\ref{eq:pn1}) for the all  possible values of the switching 
sequenc~$\delta_k$. 
Define
\begin{equation} \label{eq:pn2}
z_k(\delta_{k}) = y_{k} - \sum_{j=1}^{n_{a}}{a_{j  \delta_k}} \; y_{k-j} - \sum_{j=1}^{n_{c}}{c_{j\delta_k}} \;u_{k-j} 
\end{equation}
To identify the most probable realization of the switches, we can use the 
values of $z_k(\delta_{k})$ for each  possible active system at time $k$ 
($\delta_k = \{1,\cdots, n\}$), and determine the sequence $\delta_k$, 
$k=1,2,\ldots,N$ of maximum likelihood. 
However, for a fixed switching sequence, $z_k(\delta_k)$ is a sequence of 
correlated random variables. Even though the measurement noise is iid,  
$z_k(\delta_k)$ depends on $\eta_{k-l}\,$, $l=0,1,\ldots,n_a$ leading to a 
correlated sequence of random variables. Therefore, determining the values of 
$\delta_k\,$, $k=1,2,\ldots,N$ that lead to the highest likelihood is a complex 
combinatorial problem.
%


To circumvent this, we start by noting that $z_k(\delta_k)$ is 
independent of $z_l(\delta_l)$ if $l>k+n_a$. Therefore, if enough data is 
available, 
we can use only independent “snippets” of data of low enough length for which: 
i) maximum likelihood sequence can be easily computed and ii)  given that they 
are independent, likelihood can be computed individually for each snippet.

Hence, in the identification procedure proposed in this paper,  we  only 
consider snippets of data 
of the length $n_l$ that are separated in time by at least $n_a$ sample 
periods. More precisely, we consider snippets of data of  length $n_l$, compute 
its joint distribution as a function of the switching sequence, determine the 
maximum likelihood switches for this snippet, skip the next $n_a$ data points, 
and repeat the process until we run out of data. 

We now elaborate on this. Take  snippets of data of  length $n_l$, denoted as a 
vector 
$Z_k$ defined as:
\begin{equation}
\begin{aligned}
Z_k = &[z_k(\delta_{k}), \cdots, \, z_{k+n_l-1}(\delta_{k+n_l-1}) ]^T \\
&\forall ~k = (n_a + 1) + (n_a + n_l ) \times l , ~~\\
& \forall ~l = 0,1, 2, \cdots , \text{int}(\dfrac{N}{n_a +n_l})- (n_a +n_l + 1)
\end{aligned}
\end{equation}
where int$(\cdot)$ refers to integer part (round towards zero) of its argument.  
In this way, each snippet $Z_k$ is independent from other snippets  and 
each of these has a mutltivariable Normal distribution whose covariance matrix 
is a function of the switching sequence.
%
%
%
As a reminder, an $n_l$ dimension multivariate Normal distribution has  density 
function
\begin{align*}
	f(x)=  \dfrac{1}{(2\, \pi)^{{n_l}/2} |\Sigma|^{1/2}} \exp\{-\dfrac{1}{2}(x)^T 
	|\Sigma|^{-1} (x)\}
\end{align*}
where $\Sigma$ is the covariance matrix of dimension $n_l\times n_l$.

Note that, at each time $k$, there are $n^{n_l}$ possible switching sequences 
for the snippet $Z_k$, since 
$
\delta_{k} \in \big\{1, \cdots, \,  n \big\}.
$
Therefore, if $n_l$ is small enough, we can compute the likelihood value for 
each 
of the $n^{n_l}$ choices and take  the most likely sequence  of subsystems  
as the one 
leading to the highest likelihood. 
Hence, given the independence of $Z_k$, estimating the most likely switching 
sequence in the  used snippets can be done by solving the following 
problem
\begin{equation} \label{eq:opt1}
\begin{aligned} 
& \max_{Z_k,\delta_k}~~~\sum_{k}\log[f(Z_k)]~~~~  \\
& \text{s.t.}~~~~~~ Z_k = [z_k(\delta_{k}), \cdots, \, z_{k+n_l-1}(\delta_{k+n_l-1}) ]^T \\
&~~~~~~ ~~~\delta_{k} \in \big\{ 1, \cdots, \, n \big\}\\
&~~~~~~~~~~\forall ~k = (n_a + 1) + (n_a + n_l ) \times l , ~~\\
&~~~~~~~~~~ \forall ~l = 0,1, 2, \cdots , \text{int}(\dfrac{N}{n_a +n_l})- (n_a +n_l +1)\\
\end{aligned}
\end{equation}	
\vskip .01in  \noindent
whose optimization can be done separately  for each term of the sum. 
Therefore, its complexity is exponential in $n_l$ but linear in the number of 
snippets and can be efficiently solved if $n_l$ is ``not too large.''
\vskip.1in
\begin{Remark}
	As previously mentioned, in the above formulation, we do not use all 
	available data when computing high likelihood switchings. More precisely, 
	we only use $n_l/(n_l+n_a)$ of the data. Hence, any choice of  $n_l$ is a 
	compromise between computational complexity and fraction of the data used 
	and the ``right'' choice should be done by taking into account how many 
	measurements are available.
\end{Remark}
 
\vskip .1in 
Solving Problem~\eqref{eq:opt1}, allows us to determine how many times a 
specific "jump" occurs in this high likelihood 
sequence of switches. This can be done in the following way: 
\begin{enumerate}[Step 1.]
	\item Solve problem~\eqref{eq:opt1}. Recall that this can be done by 
	solving the problem separately for each $k$.
	\item For each $k$, let $n^{(k)}_{ij}$ be the number of times the 
	transition from system $i$ to system $j$ occurs in the maximum likelihood 
	switch sequence for snippet $k$.
	\item Compute the total number of transitions from system~$i$ to system $j$ 
	observed in all snippets
	\[
	n_{ij}=\sum_k n^{(k)}_{ij}
	\] 
\end{enumerate}

Given this high likelihood estimate of how often a transition occurs in the 
snippets, we can estimate the probability  transition matrix. This can be done 
by solving the ``traditional'' maximum likelihood problem for Markov chains
%
	\begin{equation}
	\begin{aligned} 
	& \text{maximize} 
	& & \prod_{i=1,j=1}^{n}{P_{ij}^{n_{ij}}} \\
	& \text{subject to} 
	& &    \sum_{j=1}^{n}{P_{ij}}=1, \qquad i=1,\cdots,n \\
	& & &   P_{ij}\geq 0, \qquad i=1,\cdots,n \quad j=1,\cdots,n \\       
	\end{aligned} \label{eq:opt2}
	\end{equation}
\vskip .1in 
\noindent
or equivalently, solve the equivalent convex optimization problem
\begin{equation}
\begin{aligned}
& \text{maximize} 
& &
\sum_{i=1,\; j=1}^{n}{n_{ij}\; \log({P_{ij}})} \\
& \text{subject to} 
& & \sum_{j=1}^{n}{P_{ij}}=1 \qquad i=1,\cdots,n\\
& & & P_{ij}\geq 0 \qquad i=1,\cdots,n \quad j=1,\cdots,n 
\label{eq:certain2}
\end{aligned}
\end{equation}
As the number of observations goes to infinity, the solution of this problem 
converges to the true probability transition matrix. More precisely, we have 
the following result
\vskip 0.05in
\begin{thm}
	Assume that the Markov model for switching is aperiodic and let $\hat{P}_N$ 
	and $P_{true}$ be the estimated and true transition probabilities 
	respectively. Then,
	\begin{equation}
	\lim\limits_{N \to \infty}~~\lim\limits_{\sigma^2 \to 0}~~ || \hat{P}_N - 
	P_{true}|| ~ \to 0
	\label{eq:conv}
	\end{equation}
	where $\sigma^2$ is the variance of the measurement noise.
\end{thm}

\vskip 0.1in

\noindent
\textbf{Sketch of proof:} From the results 
in~\cite{hojjatinia2018identification}, we know the estimates of  
parameters of the system converge to the true parameters as $\sigma^2 \to 0$ 
for large enough finite $N$. Hence, under mild assumptions on the subsystems, 
we 
can identify the true transitions within snippets. The fact that the Markov 
chain is not periodic, implies that, in an infinite sequence,  transitions will 
occur with equal 
probability in each snippet. This implies that, as $N \to \infty$ relative 
frequency of the transitions converges to the true transition probabilities.


\subsection{An Example}

To better illustrate the approach, we provide an example of how to do the maximum likelihood estimation of probabilities  required for identification of probability transition matrix. 
Therefore, consider the problem of identifying switching dynamics for SAR 
system  with $n=2$ subsystems of the form
\begin{equation} \label{eq:ex}
\begin{aligned}
\text{subsystem 1}: ~~& x_{k}= a_{1} \, x_{k-1} +b_{1}\, u_{k-1}  \\
\text{subsystem 2}:~~ & x_{k}= a_{2} \, x_{k-1} +b_{2}\, u_{k-1}  \\
\end{aligned}
\end{equation}
and noisy measurements
\begin{align} 
y_{k}= x_{k}+\eta_{k}  
\end{align}
where $\eta_{k}$ has zero mean Normal distribution, and $n_a = 1$. We consider snippets of data of  length $n_l = 2$, and skip $n_a = 1$ sample measurement in between the snippets of data, i.e.  two snippets of data  for subsystem 1 are like:
\begin{gather*}
\begin{aligned}
&\begin{cases}
\eta_{k} - a_1 \eta_{k-1} = y_{k} - (a_1 y_{k-1} +b_1 u_{k-1})  \\
\eta_{k+1} - a_1 \eta_{k} =  y_{k+1} - (a_1 y_{k} +b_1 u_{k})
\end{cases}
\\
&\begin{cases}
\eta_{k+3} - a_1 \eta_{k+2} = y_{k+3} - (a_1 y_{k+2} +b_1 u_{k+2})  \\
\eta_{k+4} - a_1 \eta_{k+3} =  y_{k+4} - (a_1 y_{k+3} +b_1 u_{k+3})
\end{cases}
\end{aligned}
\end{gather*}
and we have skipped this one:

$\eta_{k+2} - a_1  \eta_{k+1} = y_{k+2} - (a_1 y_{k+1} +b_1 u_{k+1})$
\vskip 0.1in
\noindent
For this example
\begin{equation*}
\begin{aligned}
Z_k = &[z_k(\delta_{k}),\, z_{k+1}(\delta_{k+1}) ]^T \\
&\forall ~k = 2 + 3\times l , ~~\\
& \forall ~l = 0,1, 2, \cdots , \text{int}(\dfrac{N}{3})- 4
\end{aligned}
\end{equation*}
and, 
\begin{align*}
z_k(\delta_{k}) \in \big\{z_k(\delta_{k} = 1),\, z_k(\delta_{k} = 2) \big\}
\end{align*}
Therefore, the set of possible active sequences can be:
\begin{equation}
\begin{aligned}
Z_k =&
\begin{cases}
[z_k(\delta_{k} = 1),\, z_{k+1}(\delta_{k+1} = 1) ]^T \\
[z_k(\delta_{k} = 1),\, z_{k+1}(\delta_{k+1} = 2) ]^T \\
 [z_k(\delta_{k} = 2),\, z_{k+1}(\delta_{k+1} = 1) ]^T \\
[z_k(\delta_{k} = 2),\, z_{k+1}(\delta_{k+1} = 2) ]^T \\
\end{cases}\\
&~~~~\forall ~k = 2 + 3\times l , ~~\\
& ~~~~\forall ~l = 0,1, 2, \cdots , \text{int}(\dfrac{N}{3})- 4
\end{aligned}
\end{equation}

\noindent
At each time $k$, there are $n^{n_l} = 4$ possible $Z_{k}$ cases. For system shown in equation (\ref{eq:ex}), $Z_{k}$ cases are as follows:
\begin{gather*}
\begin{cases}
	\eta_{k} - a_1 \eta_{k-1} = y_{k} - (a_1 y_{k-1} +b_1 u_{k-1})  \\
	\eta_{k+1} - a_1 \eta_{k} =  y_{k+1} - (a_1 y_{k} +b_1 u_{k})
\end{cases}
\\
\begin{cases}
	\eta_{k} - a_1 \eta_{k-1} = y_{k} - (a_1 y_{k-1} +b_1 u_{k-1})  \\
	\eta_{k+1} - a_2 \eta_{k} =  y_{k+1} - (a_2 y_{k} +b_2 u_{k})
\end{cases}
\\
\begin{cases}
	\eta_{k} - a_2 \eta_{k-1} = y_{k} - (a_2 y_{k-1} +b_2 u_{k-1})  \\
	\eta_{k+1} - a_1 \eta_{k} =  y_{k+1} - (a_1 y_{k} +b_1 u_{k})
\end{cases}
\\
\begin{cases}
\eta_{k} - a_2 \eta_{k-1} = y_{k} - (a_2 y_{k-1} +b_2 u_{k-1})  \\
\eta_{k+1} - a_2 \eta_{k} =  y_{k+1} - (a_2 y_{k} +b_2 u_{k})
\end{cases}
\end{gather*}

The multivariate Normal Probability at each of $n^{n_l}$ sequences $Z_k$ will 
be computed, and the one which has the maximum value of the likelihood will be 
considered as the set of active subsystems at that point. 
\section{Numerical Results} \label{result}
In this section, we will address the problem of identifying switching dynamics 
in Markovian jump systems. The values of true coefficients in this example has 
taken from the example in~\cite{hojjatinia2018identification}, which  are   
$a_{1}=0.3,~ b_{1}=1,~ a_{2}=-0.5,$ and $b_{2}=-1$. Measurement noise is 
assumed to have zero-mean Normal distribution. So, AR system with $n=2$ subsystems,  $n_a = 
1$, and $n_c = 1$ in this example are as follows:
\begin{equation} \label{eq:ex11}
\begin{aligned}
\text{subsystem 1}: ~~& x_{k}= 0.3 \, x_{k-1} +1\, u_{k-1}  \\
\text{subsystem 2}:~~ & x_{k}= -0.5 \, x_{k-1} -1\, u_{k-1}  \\
\text{noisy output}:~~ & y_{k}= x_{k} +\, \eta_{k}  \\
\end{aligned}
\end{equation}

Total number of $N=10^6$ input-output data is available.  Output is corrupted with random measurement noise which is Normal with zero mean and different values of variance.
The proposed algorithm is coded and run in Python.

{\renewcommand{\arraystretch}{1.2} 
\begin{table*}[ht]
	\caption{Identifying probability transition matrix (PTM) for different value of noise variance and different system run.}
	\label{tab:table1}
	\centering
	\begin{tabular}{||c|c|c|c|c|c||} 
		\hline \hline
		\text{experiment} &\text{True } & \text{Estimated } & \text{Normalized}& \text{} & \text{} \\
		\text{\#} &\text{PTM}& \text{PTM}& \text{Ferobenius norm} & \text{$\gamma$}& \text{$\sigma^2$}  \\
		\hline \hline 
		1	& $
		\begin{pmatrix} 
		0.1837 & 0.8163 \\
		0.3424 & 0.6576 
		\end{pmatrix} $
		 & $ \begin{pmatrix} 
		 0.2116 & 0.7884 \\
		 0.3472 & 0.6528 
		 \end{pmatrix} $
		  & 0.035810 & 0.0888& 0.01\\
		\hline
		2	&
		$\begin{pmatrix} 
		0.4286 & 0.5714 \\
		0.1412 & 0.8588 
		\end{pmatrix} $
		& $ \begin{pmatrix} 
		0.3897 & 0.6103 \\
		0.1776 & 0.8224
		\end{pmatrix} $ & 
		 0.066922 & 0.1439  &  0.03 \\
		\hline
		3	&
		$\begin{pmatrix} 
		0.1748 & 0.8252 \\
		0.4921 & 0.5079 
		\end{pmatrix} $
		& $ \begin{pmatrix} 
		0.2439 & 0.7561 \\
		0.4802 & 0.5198
		\end{pmatrix} $ & 
		 0.090075&  0.1967 &  0.05 \\
		\hline 
		4 &
		$\begin{pmatrix} 
		0.5056 & 0.4944 \\
		0.6885 & 0.3115
		\end{pmatrix} $
		& $ \begin{pmatrix} 
		0.5087 & 0.4913 \\
		0.6412 & 0.3588
		\end{pmatrix} $
		&  0.064687  &0.2273   & 0.07 \\
		\hline
		5	&
		$\begin{pmatrix} 
		0.2587 & 0.7413 \\
		0.4536 & 0.5464
		\end{pmatrix} $
		& $ \begin{pmatrix} 
		0.3200 & 0.6800 \\
		0.4474 & 0.5526
		\end{pmatrix} $
		&  0.082236 &  0.2650 & 0.09 \\
				\hline 
				6 &
				$\begin{pmatrix} 
				0.3991 & 0.6009 \\
				0.1811 & 0.8189
				\end{pmatrix} $
				& $ \begin{pmatrix} 
				0.3661 &  0.6339 \\
				0.2651& 0.7349
				\end{pmatrix} $
				& 0.115335 &  0.5223 & 0.27 \\
				\hline
				7	&
				$\begin{pmatrix} 
				0.5350 & 0.4650 \\
				0.6467 & 0.3533
				\end{pmatrix} $
				& $ \begin{pmatrix} 
				0.5180 & 0.4819 \\
				0.5788 & 0.4212
				\end{pmatrix} $
				&  0.096751 &  0.4603 & 0.29 \\
		\hline \hline
	\end{tabular}
\end{table*}
}

Noise to output ratio ($\gamma$) is defined as
\begin{equation}
\gamma = \dfrac{\max \, |\eta|}{\max \, |y|}
\end{equation}
\noindent
Simulation results for several experiments are shown in Table \ref{tab:table1}. 
For each experiment, a random probability transition matrix has been generated, 
which is shown in column 2 of the table. By using the algorithms mentioned in the 
paper, for each experiment probability transition matrix has been estimated 
from noisy measurements, which is shown in column 3 of the table. The normalized 
Frobenius norm between true and estimated values of probability transition 
matrix has been computed and shown in column 4 of the table $(
{\norm{ \hat{P} - P_{true}}_{F} }\, /\, {\norm{P_{true}}_{F}})$. 
\vskip 0.1in
For each experiment  noise to output ratio and  variance of noise are shown in 
columns 5 and 6 of the table. As we see in this table, the value of entries in 
probability transition matrix are very close to the true values, even when the 
noise variance is high with noise magnitude in average around  30\% of the 
signal magnitude.

For example, in experiment 6 the value of $\gamma =0.5223$ shows that noise to output ratio of
approximately $52\%$;
even with this very large value of corruption with measurement noise, the proposed method works well and
the normalized Frobenius norm between true and estimated values of probability transition 
matrix is only 0.1153. 
As expected and in the Table \ref{tab:table1} is shown, for smaller 
values of noise to output ratio ($\gamma$),  the estimated values for
probability transition matrix are closer to the true probability transition matrix and normalized Frobenius norm of their difference
has smaller values. 
However, with the prposed approach even for large values of 
noise to output ratio, the difference in Frobenius norm is still low.

\begin{figure}[h!]
	\begin{center}
		\centering\includegraphics[clip,width=1\columnwidth]{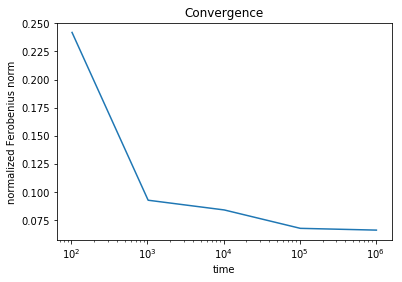}
		\caption{Convergence of estimated probability transition matrix}\label{fig:f3}
	\end{center}
\end{figure}

Figure \ref{fig:f3} demonstrates the convergence of probability transition 
matrix as number of measurements grows. This figure is based on a random 
experiment, for the values of coefficients in equation (\ref{eq:ex11}), fixed 
variance of noise  $\sigma^2 = 0.03$ and noise to output ratio $\gamma = 0.15$. 
As we see in this figure the values of normalized Frobenius norm between true 
and estimated probability transition matrix decreases, when number of 
measurement increases. As we observe in Figure \ref{fig:f3}, the value of 
difference between true and estimated probability transition matrix decreases 
from $0.2419$  at  $k = 100$ to $0.0659$  at  $k = 10^6$. It shows even for the 
case of having $15\%$ noise to output ratio, the approximated switching dynamics and 
transition probability matrix are close to the true ones.
\section{conclusion and future work } \label{conclusion} 
In this paper we have proposed a methodology for identification of switching 
dynamics in Markovian jump SAR models. Given large noisy input-output data, by 
using previously developed procedures for identification of switched system 
from large noisy data sets, 
we estimate the parameters of the noise, and then, identify the coefficients 
of each submodel. Then, by using the novel procedure presented in this 
paper for  estimation of probability transition matrix, we identify the 
switching dynamics and computed the probability transition matrix of Markov 
chain. Even for large values of measurement noise, numerical simulations show a 
low estimation error. The Frobenius norm between estimated and true probability 
transition matrix is small even in the case of large noise to output ratio. 
For future work, we can consider the problem of identifying switched ARX models 
and switching dynamics form large noisy data sets, but with the process noise. 
We  will  also test the effectiveness of the proposed  approaches in ``real'' 
applications with emphasis on estimating individual response to treatments 
aimed at improving light physical activity.

\bibliographystyle{plain}
\bibliography{arxiv_cdc19}

\begin{thebibliography}{10}

\bibitem{conroy2019}
David~E Conroy, Sarah Hojjatinia, Constantino~M Lagoa, Chih-Hsiang Yang,
  Stephanie~T Lanza, and Joshua~M Smyth.
\newblock Personalized models of physical activity responses to text message
  micro-interventions: A proof-of-concept application of control systems
  engineering methods.
\newblock {\em Psychology of Sport and Exercise}, 41:172--180, 2019.

\bibitem{harris2013algebraic}
Joe Harris.
\newblock {\em Algebraic geometry: a first course}, volume 133.
\newblock Springer Science \& Business Media, 2013.

\bibitem{HOJJATINIA201714088}
Sarah Hojjatinia, Constantino~M. Lagoa, and Fabrizio Dabbene.
\newblock A method for identification of markovian jump arx processes.
\newblock {\em IFAC-PapersOnLine}, 50(1):14088 -- 14093, 2017.
\newblock 20th IFAC World Congress.

\bibitem{hojjatinia2018identification}
Sarah Hojjatinia, Constantino~M Lagoa, and Fabrizio Dabbene.
\newblock Identification of switched arx systems from large noisy data sets.
\newblock {\em arXiv preprint arXiv:1804.07411}, 2018.

\bibitem{juloski2005bayesian}
Aleksandar~Lj Juloski, Siep Weiland, and WPMH Heemels.
\newblock A bayesian approach to identification of hybrid systems.
\newblock {\em IEEE Transactions on Automatic Control}, 50(10):1520--1533,
  2005.

\bibitem{Lagoa2017}
Constantino~M. Lagoa, David~E. Conroy, Sarah Hojjatinia, and Chih-Hsiang Yang.
\newblock Modeling subject response to interventions aimed at increasing
  physical activity: A control systems approach.
\newblock {\em Poster Presented at 5th International Conference on Ambulatory
  Monitoring of Physical Activity and Movement}, 2017.

\bibitem{nakada2005identification}
Hayato Nakada, Kiyotsugu Takaba, and Tohru Katayama.
\newblock Identification of piecewise affine systems based on statistical
  clustering technique.
\newblock {\em Automatica}, 41(5):905--913, 2005.

\bibitem{Ozay2015180}
Necmiye Ozay, Constantino Lagoa, and Mario Sznaier.
\newblock Set membership identification of switched linear systems with known
  number of subsystems.
\newblock {\em Automatica}, 51:180 -- 191, 2015.

\bibitem{ozay2012sparsification}
Necmiye Ozay, Mario Sznaier, Constantino~M Lagoa, and Octavia~I Camps.
\newblock A sparsification approach to set membership identification of
  switched affine systems.
\newblock {\em IEEE Transactions on Automatic Control}, 57(3):634--648, 2012.

\bibitem{paoletti2007identification}
Simone Paoletti, Aleksandar~Lj Juloski, Giancarlo Ferrari-Trecate, and Ren{\'e}
  Vidal.
\newblock Identification of hybrid systems a tutorial.
\newblock {\em European journal of control}, 13(2-3):242--260, 2007.

\bibitem{roll2004identification}
Jacob Roll, Alberto Bemporad, and Lennart Ljung.
\newblock Identification of piecewise affine systems via mixed-integer
  programming.
\newblock {\em Automatica}, 40(1):37--50, 2004.

\bibitem{VidalSoattoMaEtAl2003}
R.~Vidal, S.~Soatto, Yi~Ma, and S.~Sastry.
\newblock An algebraic geometric approach to the identification of a class of
  linear hybrid systems.
\newblock In {\em Proc. 42nd IEEE Conf. Decision and Control}, volume~1, pages
  167--172, December 2003.

\bibitem{vidal2005generalized}
Rene Vidal, Yi~Ma, and Shankar Sastry.
\newblock Generalized principal component analysis (gpca).
\newblock {\em IEEE transactions on pattern analysis and machine intelligence},
  27(12):1945--1959, 2005.

\bibitem{vidal2003generalized}
Rene~Esteban Vidal and Shankar Sastry.
\newblock {\em Generalized principal component analysis (gpca): an algebraic
  geometric approach to subspace clustering and motion segmentation}.
\newblock Electronics Research Laboratory, College of Engineering, University
  of California, 2003.

\end{thebibliography}
\end{document}